# Quantitative Synthesis for Concurrent Programs[⋆,⋆⋆]

Pavol Černý[1], Krishnendu Chatterjee[1], Thomas A. Henzinger[1], Arjun Radhakrishna[1], and Rohit Singh[2]

[1] IST Austria
[2] IIT Bombay

**Abstract.** We present an algorithmic method for the quantitative, performance-aware synthesis of concurrent programs. The input consists of a nondeterministic *partial program* and of a *parametric performance model*. The nondeterminism allows the programmer to omit which (if any) synchronization construct is used at a particular program location. The performance model, specified as a weighted automaton, can capture system architectures by assigning different costs to actions such as locking, context switching, and memory and cache accesses. The quantitative synthesis problem is to automatically resolve the nondeterminism of the partial program so that both correctness is guaranteed and performance is optimal. As is standard for shared memory concurrency, correctness is formalized "specification free", in particular as race freedom or deadlock freedom. For worst-case (average-case) performance, we show that the problem can be reduced to 2-player graph games (with probabilistic transitions) with quantitative objectives. While we show, using game-theoretic methods, that the synthesis problem is NEXP-complete, we present an algorithmic method and an implementation that works efficiently for concurrent programs and performance models of practical interest. We have implemented a prototype tool and used it to synthesize finite-state concurrent programs that exhibit different programming patterns, for several performance models representing different architectures.

## 1 Introduction

A promising approach to the development of correct concurrent programs is *partial program synthesis*. The goal of the approach is to allow the programmer to specify a part of her intent declaratively, by specifying which conditions, such as linearizability or deadlock freedom, need to be maintained. The synthesizer then constructs a program that satisfies the specification (see, for example, [17,16,19]).

---

[⋆] This work was partially supported by the ERC Advanced Grant QUAREM, the FWF NFN Grant S11402-N23 and S11407-N23 (RiSE), the EU NOE Grant ArtistDesign, and a Microsoft faculty fellowship.

[⋆⋆] The conference version of this paper will appear in the Proceedings of the 23rd International Conference on Computer Aided Verification (CAV), 2011.

However, quantitative considerations have been largely missing from previous frameworks for partial synthesis. In particular, there has been no possibility for a programmer to ask the synthesizer for a program that is not only correct, but also *efficient* with respect to a specific performance model. We show that providing a quantitative performance model that represents the architecture of the system on which the program is to be run can considerably improve the quality and, therefore, potential usability of synthesis.

**Motivating examples:** *Example 1.* Consider a *producer-consumer* program, where $k$ producer and $k$ consumer threads access a buffer of $n$ cells. The programmer writes a partial program implementing the procedures that access the buffer as if writing the sequential version, and specifies that at each control location a global lock or a cell-local lock can be taken. It is easy to see that there are at least two different ways of implementing correct synchronization. The first is to use a global lock, which locks the whole buffer. The second is to use cell-local locks, with each thread locking only the cell it currently accesses. The second program allows more concurrent behavior and is better in many settings. However, if the cost of locks is high (relative to the other operations), the global-locking approach is more efficient. In our experiments on a desktop machine, the global-locking implementation out-performed the cell-locking implementation by a factor of 3 in certain settings.

```
1: while(true) {
2:    lver=gver; ldata=gdata;
3:    n = choice(1..10);
4:    i = 0;
5:    while (i < n) {
6:     work(ldata); i++;
7:    }
8:    if (trylock(lock)) {
9:     if (gver==lver) {
10:       gdata = ldata;
11:       gver = lver+1;
12:       unlock(lock);
13:    } else {unlock(lock)}
14:} }
```

**Fig. 1.** Example 2

*Example 2.* Consider the program in Figure 1. It uses classic conflict resolution mechanism used for optimistic concurrency. The shared variables are `gdata`, on which some operation (given by the function `work()`) is performed repeatedly, and `gver`, the version number. Each thread has local variables `ldata` and `lver` that store local copies of the shared variables. The data is read (line 2) and operated on (line 6) without acquiring any locks. When the data is written back, the shared data is locked (line 8), and it is checked (using the version number, line 9) that no other thread has changed the data since it has been read. If the global version number has not changed, the new value is written to the shared memory (line 10), and the global version number is increased (line 11). If the global version number has changed, the whole procedure is retried. The number of operations (calls to `work`) performed optimistically without writing back to shared memory can influence the performance significantly. For approaches that perform many operations before writing back, there can be many retries and the performance can drop. On the other hand, if only a few operations are performed optimistically, the data has to be written back often, which also can lead to a performance drop. Thus, the programmer would like to leave the task of finding the optimal number of operations to be performed optimistically to the synthesizer. This is done via the choice statement (line 4).

**The partial program resolution problem.** Our aim is to synthesize concurrent programs that are both correct and optimal with respect to a performance model. The input for partial program synthesis consists of (1) a finite-state partial program, (2) a performance model, (3) a model of the scheduler, and (4) a correctness condition. A *partial program* is a finite-state concurrent program that includes nondeterministic choices which the synthesizer has to resolve. A program is *allowed* by a partial program if it can be obtained by resolving the nondeterministic choices. The second input is a *parametric performance model*, given by a weighted automaton. The automaton assigns different costs to actions such as locking, context switching, and memory and cache access. It is a flexible model that allows the assignment of costs based on past sequences of actions. For instance, if a context switch happens soon after the preceding one, then its cost might be lower due to cache effects. Similarly, we can use the automaton to specify complex cost models for memory and cache accesses. The performance model can be fixed for a particular architecture and, hence, need not be constructed separately for every partial program. The third input is the *scheduler*. Our schedulers are state-based, possibly probabilistic, models which support flexible scheduling schemes (e.g., a thread waiting for a long time may be scheduled with higher probability). In performance analysis, average-case analysis is as natural as worst-case analysis. For the average-case analysis, probabilistic schedulers are needed. The fourth input, the *correctness condition*, is a safety condition. We use "specification-free" conditions such as data-race freedom or deadlock-freedom. The output of synthesis is a program that is (a) allowed by the partial program, (b) correct with respect to the safety condition, and (c) has the best performance of all the programs satisfying (a) and (b) with respect to the performance and scheduling models.

**Quantitative games.** We show that the partial program resolution problem can be reduced to solving *imperfect information* (stochastic) graph games with quantitative (limit-average or mean-payoff) objectives. Traditionally, imperfect information graph games have been studied to answer the question of existence of general, *history-dependent* optimal strategies, in which case the problem is undecidable for quantitative objectives [8]. We show that the partial program resolution problem gives rise to the question (not studied before) whether there exist *memoryless* optimal strategies (i.e. strategies that are independent of the history) in imperfect information games. We establish that the memoryless problem for imperfect information games (as well as imperfect information stochastic games) is NP-complete, and prove that the partial program resolution problem is NEXP-complete for both average-case and worst-case performance based synthesis. We present several techniques that overcome the theoretical difficulty of NEXP-hardness in cases of programs of practical interest: (1) First, we use a lightweight static analysis technique for efficiently eliminating parts of the strategy tree. This reduces the number of strategies to be examined significantly. We then examine each strategy separately and, for each strategy, obtain a (perfect information) Markov decision process (EDP). For MDPs, efficient strategy improvement algorithms exist, and require solving Markov chains. (2)

Second, Markov chains obtained from concurrent programs typically satisfy certain progress conditions, which we exploit using a forward propagation technique together with Gaussian elimination to solve Markov chains efficiently. (3) Our third technique is to use an abstraction that preserves the value of the quantitative (limit-average) objective. An example of such an abstraction is the classical data abstraction.

**Experimental results.** In order to evaluate our synthesis algorithm, we implemented a prototype tool and applied it to four finite-state examples that illustrate basic patterns in concurrent programming. In each case, the tool automatically synthesized the optimal correct program for various performance models that represent different architectures. For the producer-consumer example, we synthesized a program where two producer and two consumer threads access a buffer with four cells. The most important parameters of the performance model are the cost $l$ of locking/unlocking and the cost $c$ of copying data from/to shared memory. If the cost $c$ is higher than $l$ (by a factor 100:1), then the fine-grained locking approach is better (by 19 percent). If the cost $l$ is equal to $c$, then the coarse-grained locking is found to perform better (by 25 percent). Referring back to the code in Figure 1, for the optimistic concurrency example and a particular performance model, the analysis found that increasing $n$ improves the performance initially, but after a small number of increments the performance started to decrease. We measured the running time of the program on a desktop machine and observed the same phenomenon.

**Related work.** Synthesis from specifications is a classical problem [6,7,15]. More recently, sketching, a technique where a partial implementation of a program is given and a correct program is generated automatically, was introduced [17] and applied to concurrent programs [16]. However, none of the above approaches consider performance-aware algorithms for sketching; they focus on qualitative synthesis without any performance measure. We know of two works where quantitative synthesis was considered. In [2,3] the authors study the synthesis of sequential systems from temporal-logic specifications. In [19,5] fixed optimization criteria (such as preferring short atomic sections or fine-grained locks) are considered. Optimizing these measures may not lead to optimal performance on all architectures. None of the cited approaches use the framework of imperfect information games, nor parametric performance models.

## 2 The Quantitative Synthesis Problem

### 2.1 Partial Programs

In this section we define threads, partial programs, programs and their semantics. We start with the definitions of guards and operations.

*Guards and operations.* Let $L$, $G$, and $I$ be finite sets of variables (representing local, global (shared), and input variables, respectively) ranging over finite domains. A *term* $t$ is either a variable in $L$, $G$, or $I$, or $t_1$ *op* $t_2$, where

$t_1$ and $t_2$ are terms and *op* is an operator. Formulas are defined by the following grammar, where $t_1$ and $t_2$ are terms and *rel* is a relational operator: $e ::= t_1 \ rel \ t_2 \ | \ e \ \wedge \ e \ | \ \neg e$. *Guards* are formulae over $L$, $G$, and $I$. *Operations* are simultaneous assignments to variables in $L \cup G$, where each variable is assigned a term over $L$, $G$, and $I$.

*Threads.* A *thread* is a tuple $\langle Q, L, G, I, \delta, \rho_0, q_0 \rangle$, with: (a) a finite set of control locations $Q$ and an initial location $q_0$; (b) $L$, $G$ and $I$ are as before; (c) an initial valuation of the variables $\rho_0$; and (d) a set $\delta$ of transition tuples of the form $(q, g, a, q')$, where $q$ and $q'$ are locations from $Q$, and $g$ and $a$ are *guards* and *operations* over variables in $L$, $G$ and $I$.

The set of locations $Sk(c)$ of a thread $c = \langle Q, L, G, I, \delta, \rho_0, q_0 \rangle$ is the subset of $Q$ containing exactly the locations where $\delta$ is non-deterministic, i.e., locations where there exists a valuation of variables in $L$, $G$ and $I$, for which there are multiple transitions whose guards evaluate to true.

*Partial programs and programs.* A *partial program* $M$ is a set of threads that have the same set of global variables $G$ and whose initial valuation of variables in $G$ is the same. Informally, the semantics of a partial program is a parallel composition of threads. The set $G$ represents the shared memory. A *program* is a partial program, in which the set $Sk(c)$ of each thread $c$ is empty. A program $P$ is *allowed* by a partial program $M$ if it can be obtained by removing the outgoing transitions from sketch locations of all the threads of $M$, so that the transition function of each thread becomes deterministic.

The guarded operations allow us to model basic concurrency constructs such as locks (for example, as variables in $G$ and locking/unlocking is done using guarded operations) and compare-and-set. As partial program defined as a collection of fixed threads, thread creation is not supported.

*Semantics.* A *transition system* is a tuple $\langle S, A, \Delta, s_0 \rangle$ where $S$ is a finite set of states, $A$ is a finite set of actions, $\Delta \subseteq S \times A \times S$ is a set of transitions and $s_0$ is the initial state. The semantics of a partial program $M$ is given in terms of a transition system (denoted as $\mathsf{Tr}(M)$). Given a partial program $M$ with $n$ threads, let $\mathcal{C} = \{1, \ldots, n\}$ represent the set of threads of $M$.

- *State space.* Each state $s \in S$ of $\mathsf{Tr}(M)$ contains input and local variable valuations and locations for each thread in $\mathcal{C}$, and a valuation of the global variables. In addition, it contains a value $\sigma \in \mathcal{C} \cup \{*\}$, indicating which (if any) thread is currently scheduled. The initial state contains the initial locations of all threads and the initial valuations $\rho_0$, and the value $*$ indicating that no thread is currently scheduled.
- *Transition.* The transition function $\Delta$ defines interleaving semantics for partial programs. There are two types of transitions: thread transitions, that model one step of a scheduled thread, and environment transitions, that model input from the environment and the scheduler. For every $c \in \mathcal{C}$, there exists a thread transition labeled $c$ from a state $s$ to a state $s'$ if and only if there exists a transition $(q, g, a, q')$ of $c$ such that (i) $\sigma = c$ in $s$ (indicating that $c$ is scheduled) and $\sigma = *$ in $s'$, (ii) the location of $c$ is $q$ in $s$ and $q'$ in $s'$, (iii) the guard $g$ evaluates to true in $s$, and (iv) the valuation of local variables of $c$ and global variables in $s$ is obtained from the valuation of variables in $s'$ by performing the operation $a$. There is an environment transition labeled $c$ from state $s$ to state $s'$ in $\mathsf{Tr}(M)$ if and only if (i) the value $\sigma$ in $s$ is $*$ and the value $\sigma$ in $s'$ is $c$ and (ii) the valuations of variables in $s$ and $s'$ differ only in input variables of the thread $c$.

### 2.2 The performance model

We define a flexible and expressive performance model via a weighted automaton that specifies costs of actions. A *performance automaton* $W$ is a tuple $W = (Q_W, \Sigma, \delta, q_0, \gamma)$, where $Q_W$ is a set of states, $\Sigma$ is a finite alphabet, $\delta : Q_W \times \Sigma \to Q_W$ is a transition relation, $q_0$ is an initial location and $\gamma$ is a cost function $\gamma : Q_W \times \Sigma \times Q_W \to \mathbb{Q}$. The labels in $\Sigma$ represent (concurrency related) actions that incur costs, while the values of the function $\gamma$ specify these costs. The symbols in $\Sigma$ are matched with the actions performed by the system to which the performance measures are applied. A special symbol $ot \in \Sigma$ denotes that none of the tracked actions occurred. The costs that can be specified in this way include for example the cost of locking, the access to the (shared) main memory or the cost of context switches.

An example specification that uses the costs mentioned above is the automaton $W$ in Figure 2. The automaton describes the costs for locking ($l$), context switching ($cs$), and main memory access ($m$). Specifying the costs via a weighted automaton is more general than only specifying a list of costs. For example, automaton based specification enables us to model a cache, and the cost of reading from a cache versus reading from the main memory, as shown in Figure 8 in Section 5. Note that the performance model can be fixed for a particular architecture. This eliminates the need to construct a performance model for the synthesis of each partial program.

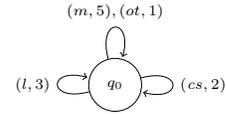

**Fig. 2.** Perf. aut.

### 2.3 The partial program resolution problem

*Weighted probabilistic transition system (WPTS).* A *probabilistic transition system* (PTS) is a generalization of a transition system with a probabilistic transition function. Formally, let $\mathcal{D}(S)$ denote the set of probability distributions over $S$. A PTS consists of a tuple $\langle S, A, \Delta, s_0 \rangle$ where $S$, $A$, $s_0$ are defined as for transition systems, and $\Delta : S \times A \to \mathcal{D}(S)$ is probabilistic, i.e., given a state and an action, it returns a probability distribution over successor states. A WPTS consists of a PTS and a weight function $\gamma : S \times A \times S \to \mathbb{Q} \cup \{\infty\}$ that assigns costs to transitions. An *execution* of a weighted probabilistic transition system is an infinite sequence of the form $(s_0 a_0 s_1 a_2 \ldots)$ where $s_i \in S$, $a_i \in A$, and $\Delta(s_i, a_i)(s_{i+1}) > 0$, for all $i \geq 0$. We now define boolean and quantitative objectives for WPTS.

*Safety objectives.* A *safety objective* $\text{Safety}_B$ is defined by a set $B$ of "bad" states and requires that states in $B$ are never present in an execution. An execution $e = (s_0 a_0 s_1 a_2 \ldots)$ is *safe* (denoted by $e \in \text{Safety}_B$) if $s_i \notin B$, for all $i \geq 0$.

*Limit-average and limit-average safety objectives.* The *limit-average* objective assigns a real-valued quantity to every infinite execution $e$. We have $\text{LimAvg}_\gamma(s_0 a_0 s_1 a_1 s_2 \ldots) = \limsup_{n \to \infty} \frac{1}{n} \sum_{i=0}^{n} \gamma((s_i, a, s_{i+1}))$ if there are no infinite cost transitions, and $\infty$ otherwise. The *limit-average safety* objective (defined by $\gamma$ and $B$) is a *lexicographic* combination of a safety and a limit-average objective: $\text{LimAvg}_\gamma^B(e) = \text{LimAvg}_\gamma(e)$, if $e \in \text{Safety}_B$, and $\infty$ otherwise. Limit-average safety objectives can be reduced to limit-average objectives by making states in $B$ absorbing (states with only self-loop transitions) and assigning the self-loop transitions the weight $\infty$.

*Value of WPTS.* Given a WPTS $T$ with weight function $\gamma$, a *policy pf* : $(S \times A)^* \times S \to A$ is a function that given a sequence of states and actions chooses an action. A policy $pf$ defines a unique probability measure on the executions and let $E^{pf}(\cdot)$ be the associated expectation measure. Given a WPTS $T$ with weight function $\gamma$, and a policy $pf$, the value $\text{Val}(T, \gamma, \text{Safety}_B, pf)$ is the expected value $E^{pf}(\text{LimAvg}_\gamma^B)$ of the limit-average safety objective. The *value* of the WPTS is the supremum value over all policy functions, i.e., $\text{Val}(T, \gamma, \text{Safety}_B) = \sup_{pf} \text{Val}(T, \gamma, \text{Safety}_B, pf)$.

*Schedulers.* A *scheduler* has a finite set of internal memory states $Q_{\text{Sch}}$. At each step, it considers all the active threads and chooses one either (i) nondeterministically (nondeterministic schedulers) or (ii) according to a probability distribution (probabilistic schedulers), which depends on the current internal memory state.

*Composing a program with a scheduler and a performance model.* In order to evaluate the performance of a program, we need to take into account the scheduler and the performance model. Given a program $P$, a scheduler Sch, and a performance model $W$, we construct a WPTS, denoted $\text{Tr}(P, \text{Sch}, W)$, with a weight function $\gamma$ as follows. A state $s$ of $\text{Tr}(P, \text{Sch}, W)$ is composed of a state of the transition system of $P$ ($\text{Tr}(P)$), a state of the scheduler Sch and a state of the performance model $W$. The transition function matches environment transitions of $\text{Tr}(P)$ with the scheduler transitions (which allows the scheduler to schedule threads) and it matches thread transitions with the performance model transitions. The weight function $\gamma$ assigns costs to edges as given by the weighted automaton $W$. Furthermore, as the limit average objective is defined only for infinite executions, for terminating safe executions of the program we add an edge back to the initial state. The value of the limit average objective function of the infinite execution is the same as the average over the original finite execution. Note that the performance model can specify a locking cost, while the program model does not specifically mention locking. We thus need to specifically designate which shared memory variables are used for locking.

*Correctness.* We restrict our attention to safety conditions for correctness. We illustrate how various correctness conditions for concurrent programs can be modelled as Safety objectives: (a) *Data-race freedom.* Data-races occur when two or more threads access the same shared memory location and one of the accesses is a write access. We can check for absence of data-races by denoting as unsafe states those in which there exist two enabled transitions (with at least one being a write) accessing a particular shared variable, from different threads. (b) *Deadlock freedom.* One of the major problems of synchronizing programs using blocking primitives such as locks is that deadlocks may arise. A deadlock occurs when two (or more) threads are waiting for each other to finish an operation. Deadlock-freedom is a safety property. The unsafe states are those where there exists two or more threads with each one waiting for a resource held by the next one.

*Value of a program and of a partial program.* For $P$, Sch, $W$ as before and Safety$_B$ is a safety objective, we define the value of the program using the composition of $P$, Sch and $W$ as: $ValProg(P, \text{Sch}, W, \text{Safety}_B) = Val(\text{Tr}(P, \text{Sch}, W), \gamma, \text{Safety}_B)$. For be a partial program $M$, let $\mathcal{P}$ be the set of all allowed programs. The value of $M$, $ValParProg(M, \text{Sch}, W, \text{Safety}_B) = \min_{P \in \mathcal{P}} ValProg(P, \text{Sch}, W, \text{Safety}_B)$.

*Partial Program resolution problem.* The central technical questions we address are as follows: (1) The *partial program resolution optimization problem* consists of a partial program $M$, a scheduler Sch, a performance model $W$ and a safety condition Safety$_B$, and asks for a program $P$ allowed by the partial program $M$ such that the value $ValProg(P, \text{Sch}, W, \text{Safety}_B)$ is minimized. Informally, we have: (i) if the value $ValParProg(M, \text{Sch}, W, \text{Safety}_B)$ is $\infty$, then no safe program exists; (ii) if it is finite, then the answer is the optimal safe program, i.e., a correct program that is optimal with respect to the performance model. The *partial program resolution decision problem* consists of the above inputs and a rational threshold $\lambda$, and asks whether $ValParProg(M, \text{Sch}, W, \text{Safety}_B) \leq \lambda$.

## 3 Quantitative Games on Graphs

Games for synthesis of controllers and sequential systems from specifications have been well studied in literature. We show how the partial program resolution problems can be reduced to quantitative imperfect information games on graphs. We also show that the arising technical questions on game graphs is different from the classical problems on quantitative graph games.

### 3.1 Imperfect information games for partial program resolution

An *imperfect information stochastic game graph* is a tuple $\mathcal{G} = \langle S, A, \text{En}, \Delta, (S_1, S_2), O, \eta, s_0 \rangle$, where $S$ is a finite set of states, $A$ is a finite set of actions, $\text{En} : S \to 2^A \setminus \emptyset$ is a function that maps every state $s$ to the non-empty set of actions enabled at $s$, and $s_0$ is an initial state. The transition function $\Delta : S \times A \to \mathcal{D}(S)$ is a probabilistic function which maps a state $s$ and an enabled action $a$ to the probability distribution $\Delta(s, a)$ over the successor states. The sets $(S_1, S_2)$ define a partition of $S$ into Player-1 and Player-2 states, respectively; and the function $\eta : S \to O$ maps every state to an observation from the finite observation set $O$. We refer to these as ImpIn $2\frac{1}{2}$-player game graphs: ImpIn for imperfect information, 2 for the two players and $\frac{1}{2}$ for the probabilistic transitions.

*Special cases.* We also consider ImpIn 2-player games (no randomness), perfect information games (no partial information), MDPs (only one enabled action for Player 1 states) and Markov chains (only one enabled action for all states) as special cases of ImpIn $2\frac{1}{2}$-player games. For full definitions, see [4].

The informal semantics for an imperfect information game is as follows: the game starts with a token being placed on the initial state. In each step, Player 2 can observe the exact state $s$ in which the token is placed whereas, Player 1 can observe only $\eta(s)$. If the token is in $S_1$ (resp. $S_2$), Player 1 (resp. Player 2) chooses an action $a$ enabled in $s$. The token is then moved to a successor of $s$ based on the distribution $\Delta(s, a)$.

A strategy for Player 1 (Player 2) is a "recipe" that chooses an action for her based on the history of observations (states). Memoryless Player 1 (Player 2) strategies are those which choose an action based only on the current observation (state). We denote the set of Player 1 and Player 2 strategies by $\Sigma$ and $\Gamma$, respectively, and the set of Player 1 and Player 2 memoryless strategies by $\Sigma^M$ and $\Gamma^M$, respectively.

*Probability space and objectives* Given a pair of Player 1 and Player 2 strategies $(\sigma, \tau)$, it is possible to define a unique probability measure $\Pr^{\sigma,\tau}(\cdot)$ over the set of paths of the game graph. For details, refer to any standard work on $2\frac{1}{2}$-player stochastic games (for example, [18]).

In a graph game, the goal of Player 1, i.e., the *objective* is given as a boolean or quantitative function from paths in the game graph to either $\{0, 1\}$, or $\mathbb{R}$. We consider only the LimAvg-Safety objectives defined in Section 2. Player 1 tries to maximize the expected value of the objective. The *value* of a Player 1 strategy $\sigma$ is defined as $ValGame(f, \mathcal{G}, \sigma) = \sup_{\tau \in \Gamma} \mathbb{E}^{\sigma,\tau}[f]$ and the value of the game is defined as $ValGame(f, \mathcal{G}) = \inf_{\sigma \in \Sigma} ValGame(f, \mathcal{G}, \sigma)$.

For a more detailed exposition on ImpIn $2\frac{1}{2}$-player graph games and the formal definition of strategies, objectives, and values, see [4].

*Decision problems.* Given a game graph $\mathcal{G}$, an objective $f$ and a rational threshold $q \in \mathbb{Q}$, the general decision problem (resp. memoryless decision problem) asks if there is a Player 1 strategy (resp. memoryless strategy) $\sigma$ with $ValGame(f, \mathcal{G}, \sigma) \leq q$. Similarly, the value problem (memoryless value problem) is to compute $\inf_{\sigma \in \Sigma} ValGame(f, \mathcal{G}, \sigma)$ ($\min_{\sigma \in \Sigma^M} ValGame(f, \mathcal{G}, \sigma)$ resp.). Traditional game theory study always considers the general decision problem which is undecidable for limit-average objectives [8] in imperfect information games.

**Theorem 1.** *[8] The decision problems for* LimAvg *and* LimAvg-Safety *objectives are undecidable for* ImpIn $2\frac{1}{2}$- *and* ImpIn *2-player game graphs.*

However, we show here that the partial program resolution problems reduce to the memoryless decision problem for imperfect information games.

**Theorem 2.** *Given a partial program $M$, a scheduler* Sch, *a performance model $W$, and a correctness condition $\phi$, we construct an exponential-size* ImpIn $2\frac{1}{2}$-*player game graph $\mathcal{G}_M^p$ with a* LimAvg-Safety *objective such that the memoryless value of $\mathcal{G}_M^p$ is equal to* $ValParProg(M, \text{Sch}, W, \text{Safety})$.

*Proof.* The proof relies on the construction of an imperfect information game graph, denoted $\mathcal{G}(M, \text{Sch}, W)$, in which fixing a memoryless strategy $\sigma$ for Player 1 yields a WPTS $\text{Tr}(P_\sigma, \text{Sch}, W)$ with weight function $\gamma$ that corresponds to the product of a program $P_\sigma$ allowed by the partial program $M$, composed with the scheduler Sch and the performance model $W$. The construction of this game graph is similar to the construction of the product of a program, scheduler and performance model, but with a partial program replacing the program. Due to the nondeterministic transition function of the partial program, there will exist extra nondeterministic choices in the WPTS (in addition to the choice of inputs). This nondeterminism is resolved by Player 1 choices and the nondeterminism due to input (and possibly scheduling) is resolved by Player 2 choices. We refer to this game as the *program resolution game*.

The crucial point of the construction is the observations, i.e., the information about the state that is visible to Player 1. Since Player 1 is to resolve the nondeterminism from the partial program, he is allowed only to observe the scheduled thread and its current location. He may choose a set of transitions, from that location, such that only one of the set is enabled for any valuation of the variables. The formal description of the reduction of partial program resolution to imperfect information games is as follows.

- *State space.* Analogous to the construction of $\text{Tr}(P, \text{Sch}, W)$, a state in the state space of $\mathcal{G}(M, \text{Sch}, W)$ is a tuple $(s, q_{\text{Sch}}, q_W)$ where $s$, $q_{\text{Sch}}$ and $q_W$ are states of $\text{Tr}(M)$, Sch and $W$, respectively.
- *Player-1 and Player-2 partition.* The state is a Player 1 state if $s$ is labelled with a scheduled thread, and a Player 2 state if $s$ has no thread scheduled and is labelled with a $*$.
- *Observation and observation mapping.* The set of observations $O$ is the set of locations from all the threads of $M$ along with a $\bot$ element, i.e., $O = \{\bot\} \cup \{(t, q) | t$ is a thread of $M$ and $q$ is a partial program location of $t\}$. All Player 2 states are mapped to $\bot$ by $\eta$. Player 1 states with thread $t$ scheduled and thread $t$ in location $q$ are mapped to $(t, q)$ by $\eta$.
- *Enabled actions and transitions.* Suppose $(s, q_{\text{Sch}}, q_W)$ is a Player 1 state with $\eta((s, q_{\text{Sch}}, q_W)) = (t, q)$. Any action $a$ enabled in this state is a set of transitions of thread $t$ from state $q$ such that only one of them is enabled for any valuation of local, global and input variables. On choosing action $a$ in $(s, q_{\text{Sch}}, q_W)$, the control moves to the state $(s', q'_{\text{Sch}}, q'_W)$ where $s'$ is the state obtained by executing the unique enabled transition from $a$ in $s$. The states $q'_{\text{Sch}}$ and $q'_W$ are as in $\text{Tr}(P, \text{Sch}, W)$. The set of Player 2 actions and transitions are as in $\text{Tr}(P, \text{Sch}, W)$.
- *Initial state.* The initial state of $\mathcal{G}(M, \text{Sch}, W)$ is the tuple of initial states of $M$, Sch and $W$.

To complete the proof, we show that given a memoryless Player 1 strategy $\sigma$, there exists a program $P_\sigma$ allowed by $M$ such that $\text{Tr}(P_\sigma, \text{Sch}, W)$ corresponds to the MDP obtained by fixing $\sigma$ in $\mathcal{G}(M, \text{Sch}, W)$ and vice-versa.

Given a program $P_\sigma$ allowed by the partial program, we construct a memoryless $\sigma$ as follows: $\sigma((t, q))$ is the action consisting of the set of transitions from location $q$ in thread $t$ in $P_\sigma$. As $P_\sigma$ is deterministic, only one of them will be enabled for a valuation of the variables. Similarly, given a memoryless Player 1 strategy, we construct $P_\sigma$ by preserving only those transitions from location $q$ of thread $t$ which are present in $\sigma((t,q))$. From the above construction we conclude the desired correspondence. □

### 3.2 Complexity of ImpIn Games and partial-program resolution

We establish complexity bounds for the relevant memoryless decision problems and use them to establish upper bounds for the partial program resolution problem. We also show a matching lower bound. First, we state a theorem on complexity of MDPs.

**Theorem 3.** *[9] The memoryless decision problem for* LimAvg-Safety *objectives can be solved in polynomial time for MDPs.*

**Theorem 4.** *The memoryless decision problems for* Safety, LimAvg, *and* LimAvg-Safety *objectives are* NP-*complete for* ImpIn $2\frac{1}{2}$- *and* ImpIn 2-*player game graphs.*

For the lower bound we show a reduction from 3SAT problem and for the upper bound we use memoryless strategies as polynomial witness and Theorem 3 for polynomial time verification procedure. We prove below in two lemmas to establish Theorem 4.

**Lemma 1.** *The memoryless decision problem for* ImpIn 2-*player game graphs with* Safety *and* LimAvg *objectives are* NP-*hard.*

*Proof.* We first show NP-hardness for safety objectives.

*(*NP-*hardness).* We will show that the memoryless decision problem for ImpIn 2-player safety game is NP-hard by reducing the 3-SAT problem. Given a 3-SAT formula $\Phi$ over variables $x_1, x_2, \ldots x_N$, with clauses $C_1, C_2, \ldots C_K$, we construct an imperfect information game graph with $N+1$ observations and $3K+2$ states such that Player 1 has a memoryless winning strategy from the initial state if and only if $\Phi$ is satisfiable. The construction is described below:

- The states of the game graph are $\{init\} \cup \{s_{i,j} \mid i \in [1,K] \land j \in \{1,2,3\}\} \cup \{bad\}$.
- The observations and the observation mapping are as follows: *init* and *bad* are mapped with observation 0, and $s_{i,j}$ is mapped with observation $k$ if the $j^{th}$ variable of the $C_i$ clause is $x_k$ or $\neg x_k$.
- *init* and *bad* are Player 2 states and all other states are Player 1 states.
- The actions and transition function of the game graph are as follows:
    1. For all $i \in [0, K]$, there is a transition from *init* to $s_{i,1}$ on the action $\bot$.
    2. If the $j^{th}$ literal of clause $C_i$ is $x_k$, then there are two actions enabled (*true* and *false*) and there is a transition from $s_{i,j}$ to *init* on *true* and to $s_{i+1,j}$ on *false* (for $j \in \{1,2\}$). For $j=3$, the transition on *true* leads to *init* and the transition on *false* leads to *bad*.

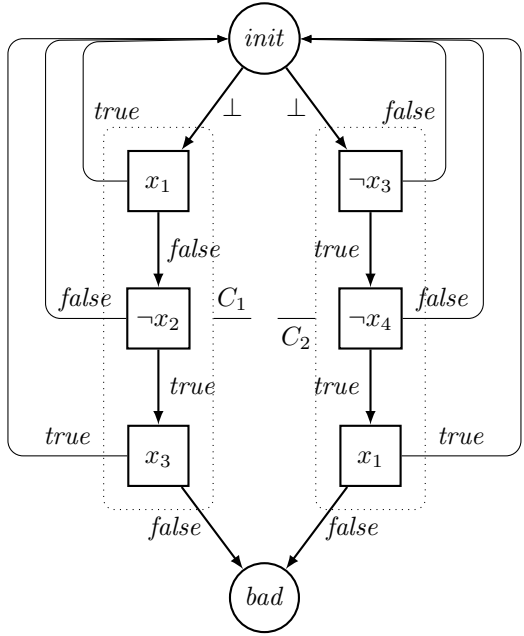

**Fig. 3.** 3-SAT to memoryless imperfect information Safety games

3. If the $j^{th}$ literal of clause $C_i$ is $\neg x_k$, there is a transition from $s_{i,j}$ to *init* on *false* and to $s_{i+1,j}$ on *true* (for $j \in \{1,2\}$). For $j = 3$, the transition on *false* leads to *init* and the edge on *true* leads to *bad*.
– The objective for Player 1 is to avoid reaching *bad* and the objective for Player 2 is to reach *bad*.

Intuitively, Player 2 chooses a clause $C_i$ in the initial state *init*. Player 1 then plays according to her memoryless strategy from each of the states $s_{i,j}$. If the action $a \in \{true, false\}$ chosen in $s_{i,j}$ makes the literal at position $j$ in clause $C_i$ true, control goes back to *init*. Otherwise, the control goes to the next $s_{i,j+1}$. If the choices at all three $s_{i,j}$'s make the corresponding literal false, the control goes to *bad*. The game graph structure is illustrated in Figure 3.

Given a truth value assignment of $x_i$'s such that $\Phi$ is satisfied, we can construct a memoryless strategy of Player 1 which chooses the action at observation $i$ same as the valuation of $x_i$, and the memoryless strategy is winning for Player 1. In every triple of $s_{i,j}$'s at least one of the edges dictated by this strategy lead to *init*. If that were not the case, the corresponding clause would not have been satisfied. Given a winning memoryless strategy $\tau_1$, the valuation of $x_i$'s which assigns the $\tau_1(i)$ to $x_i$ satisfies each clause $C_k$ in $\Phi$. This follows from a similar argument as above. Hence the hardness result follows.

The above reduction is slightly modified to show that the LimAvg memoryless decision problem is also NP-hard. This can be done by adding a self loop on state *bad* with weight 1 and attaching the weight 0 to all other edges. Now, Player 1 can obtain a value less than 1 if and only if she has a memoryless winning strategy in the Safety game. The desired result follows. □

**Lemma 2.** *The memoryless decision problem for* LimAvg-Safety *objectives for* ImpIn $2\frac{1}{2}$*-player game graphs is in* NP.

*Proof.* Given a memoryless winning strategy for a Player 1 in a ImpIn $2\frac{1}{2}$-player game graph, the verification problem is equivalent to solving for the same objective on the MDP obtained by fixing the strategy for Player 1. Hence the memoryless strategy is the polynomial witness, and Theorem 3 provides the polynomial time verification procedure to prove the desired result. □

Lemma 1 and Lemma 2 gives us Theorem 4.

*Remark 1.* The NP-completeness of the memoryless decision problems rules out the existence of the classical strategy improvement algorithms as their existence implies existence of randomized sub-exponential time algorithms (using the techniques of [1]), and hence a strategy improvement algorithm would imply a randomized sub-exponential algorithm for an NP-complete problem.

**Theorem 5.** *The partial-program resolution decision problem is* NEXP-*complete for both nondeterministic and probabilistic schedulers.*

*Proof.* (a) The NEXP upper bound follows by an exponential reduction to the ImpIn games' memoryless decision problem (Theorem 2), and by Theorem 4.

(b) We reduce the NEXP-hard problem *succinct 3-SAT* (see [14]) to the partial program resolution problem to show NEXP-hardness. The idea is to construct a two thread partial program (shown in Figure 4) where Thread 1 chooses a clause from the formula and Thread 2 will determine the literals in the clause and then, enters an error state if the clause is not satisfied.

```
GLOBALS: var i;

THREAD 1:
while (true)
    i = (i + 1) mod N;

THREAD 2:
choice: {
    val[v_1] = true;
    val[v_1] = false;
}
...

while (true)
    l1 = compute_Q(i,1);
    l2 = compute_Q(i,2);
    l3 = compute_Q(i,3);
    if(not (val[l1] ∨
            val[l2] ∨
            val[l3]))
        assert(false);
```

**Fig. 4.** The reduction of succinct 3-SAT to partial program resolution

Given an instance of succinct 3-SAT over variables $v_1, \ldots v_M$, i.e., a circuit $\mathcal{Q}$ which takes pairs $(i,j)$ and returns the $j^{th}$ literal in the $i^{th}$ clause. Thread 1 just changes the global variable $i$, looping through all clause indices.

Thread 2 will first non-deterministically choose a valuation $\mathcal{V}$ for all literals. It then does the following repeatedly: (a) Read global $i$, (b) Compute the $i^{th}$ clause by solving the circuit value problem for $\mathcal{Q}$ with $(i,1)$, $(i,2)$ and $(i,3)$ as inputs. This can be done in polynomial time. (c) If the $i^{th}$ clause is not satisfied with the valuation $\mathcal{V}$, it goes to an error state.

To show the validity of the reduction: (i) Given a satisfying valuation for $\mathcal{Q}$, choosing that valuation in the first steps of Thread 2 will obviously generate a safe program. (ii) Otherwise, for every valuation $\mathcal{V}$ chosen in the partial program, there exists a clause (say $k$) which is not satisfied. We let Thread 1 run till $i$ becomes equal to $k$ and then let Thread 2 run. The program will obviously enter the error state. Note that the result is independent of schedulers (non-deterministic or probabilistic), and performance models (as it uses only safety objectives). □

## 4 Practical Solutions for Partial-Program Resolution

**Algorithm 1** Strategy Elimination
**Input:** $M$: partial program;
$W$: performance model;
Sch: scheduler;
Safety: safety condition
**Output:** *Candidates*: Strategies
$StrategySet \leftarrow \texttt{CompleteTree}(M)$
{A complete strategy tree}
$Candidates \leftarrow \emptyset$
**while** $StrategySet \neq \emptyset$ **do**
  Choose *Tree* from *StrategySet*
  $\sigma \leftarrow \texttt{Root}(Tree)$
  **if** $\texttt{PartialCheck}(\sigma, \text{Safety})$ **then**
    $StrategySet =$
      $StrategySet \cup \texttt{children}(Tree)$
  **if** *Tree* is singleton **then**
    $Candidates = Candidates \cup \{\sigma\}$
**return** *Candidates*

We present practical solutions for the computationally hard (NEXP-complete) partial-program resolution problem.

**Strategy elimination.** We present the general strategy enumeration scheme for partial program resolution. We first introduce the notions of a partial strategy and strategy tree.

*Partial strategy and strategy trees.* A *partial memoryless strategy* for Player 1 is a partial function from observations to actions. A *strategy tree* is a finite branching tree labelled with partial memoryless strategies of Player 1 such that: (a) Every leaf node is labelled with a complete strategy; (b) Every node is labelled with a unique partial strategy; and (c) For any parent-child node pair, the label of the child ($\sigma_2$) is a proper extension of the label of parent ($\sigma_1$), i.e., $\sigma_1(o) = \sigma_2(o)$ when both are defined and the domain of $\sigma_2$ a proper superset of $\sigma_1$. A complete strategy tree is one where all Player 1 memoryless strategies are present as labels.

In the strategy enumeration scheme, we maintain a set of candidate strategy trees and check each one for partial correctness. If the root label of the tree fails the partial correctness check, then remove the whole tree from the set. Otherwise, we replace it with the children of the root node. The initial set is a single complete strategy tree. In practice, the choice of this tree can be instrumental in the efficiency of partial correctness checks. Trees which first fix the choices that help the partial correctness check to identify an incorrect partial strategy are more useful. The partial program resolution scheme is shown in Algorithm 1.

The `PartialCheck` function checks for the partial correctness of partial strategies, and returns "Incorrect" if it is able to prove that all strategies compatible with the input are unsafe, or it returns "Don't know". In practice, for the partial correctness checks the following steps can be used: (a) checking of lock discipline to prevent deadlocks; and (b) simulation of the partial program on small inputs;

**Algorithm 2** Synthesis Scheme
---
**Input:** $M$: partial program;
   $W$: performance model;
   Sch: a scheduler;
   Safety: a safety condition
**Output:** $P$: correct program or $\bot$
   $Candidates \leftarrow \texttt{StrategyElimination}(M, \text{Sch}, W, \text{Safety})$
   $StrategyValues \leftarrow \emptyset$
   **while** $Candidates \neq \emptyset$ **do**
      Pick $\sigma$ from $Candidates$
      $\mathcal{G}_\sigma \leftarrow \mathcal{G}(M, \text{Sch}, W)$ with $\sigma$ fixed
      $\mathcal{G}_\sigma^* \leftarrow \texttt{Abstract}(\mathcal{G}_\sigma)$
      $Valid \leftarrow \texttt{SoftwareModelCheck}(\mathcal{G}_\sigma^*, \text{Safety})$
      **if** $Valid$ **then**
         $Value \leftarrow \texttt{SolveMDP}(\mathcal{G}_\sigma^*)$
         $StrategyValues \leftarrow StrategyValues \cup \{\sigma \mapsto Value\}$
   **if** $StrategyValues = \emptyset$ **then**
      **return** $\bot$
   **else**
      $OptimalStrategy = \texttt{minimum}(StrategyValues)$
      **return** $M$ with $OptimalStrategy$ strategy fixed
---

The result of the scheme is a set of candidate strategies for which we evaluate full correctness and compute the value. The algorithm is shown in Algorithm 2. In the algorithm, the procedures `SoftwareModelCheck`, `Abstract` and `SolveMDP` are of special interest. The procedure `Abstract` abstracts an MDP preserving the LimAvg-Safety properties as described in the following paragraphs. The `SolveMDP` procedure uses the optimizations described below to compute the LimAvg value of an MDP efficiently. The Safety conditions are checked by `SoftwareModelCheck` procedure. It might not explicitly construct the states of the MDP, but may use symbolic techniques to check the Safety property on the MDP. It is likely that further abstraction of the MDP may be possible during this procedure as we need abstractions which preserve Safety, and $\mathcal{G}_\sigma^*$ is abstracted to preserve both Safety and LimAvg values.

**Evaluation of a memoryless strategy.** Fixing a memoryless Player 1 strategy in a ImpIn $2\frac{1}{2}$-player game for partial program resolution gives us (i) a non-deterministic transition system in the case of a non-deterministic scheduler, or (ii) an MDP in case of probabilistic schedulers. These are perfect-information games and hence, can be solved efficiently. In case (i), we use a standard min-mean cycle algorithm (for example, [12]) to find the value of the strategy . In case (ii), we focus on solving Markov chains with limit-average objectives efficiently. Markov chains arise from MDPs due to two reasons: (1) In many cases, program input can be abstracted away using data abstraction and the problem is reduced to solving a LimAvg Markov Chain. (2) The most efficient algorithm for LimAvg MDPs is the strategy improvement algorithm [9], and each step of the algorithm involves solving a Markov chain (for standard techniques, see [9]).

In practice, a large fraction of concurrent programs are designed to ensure progress condition called *lock-freedom* [10]. Lock-freedom ensures that some thread always makes progress in a finite number of steps. This leads to Markov chains with a directed-acyclic tree like structure with only few cycles introduced to eliminate finite executions as mentioned in Section 2. We present a *forward propagation* technique to compute stationary probabilities for these Markov chains. Computing the stationary distribution for a Markov chain involves solving a set of linear equalities using Gaussian elimination. For Markov chains that satisfy the special structure, we speed up the process by eliminating variables in the tree by forward propagating the root variable. Using this technique, we were able to handle the special Markov chains of up to 100,000 states in a few seconds in the experiments.

**Quantitative probabilistic abstraction.** To improve the performance of the synthesis, we use standard abstraction techniques. However, for the partial program resolution problem we require abstraction that also preserves quantitative objectives such as LimAvg and LimAvg-Safety. We show that an extension of probabilistic bisimilarity [13] with a condition for weight function preserves the quantitative objectives.

*Quantitative probabilistic bisimilarity.* A binary equivalence relation $\equiv$ on the states of a MDP is a *quantitative probabilistic bisimilarity* relation if (a) $s \equiv s'$ iff $s$ and $s'$ are both safe or both unsafe; (b) $\forall s \equiv s', a \in A : \sum_{t \in C} \Delta(s, a)(t) = \sum_{t \in C} \Delta(s', a)(t)$ where $C$ is an equivalence class of $\equiv$; and (c) $s \equiv s' \wedge t \equiv t' \implies \gamma(s, a, s') = \gamma(t, a, t')$. The states $s$ and $s'$ are *quantitative probabilistic bisimilar* if $s \equiv s'$.

A *quotient* of an MDP $\mathcal{G}$ under quantitative probabilistic bisimilarity relation $\equiv$ is an MDP ($\mathcal{G}/_\equiv$) where the states are the equivalence classes of $\equiv$ and: (i) $\gamma(C, a, C') = \gamma(s, a, s')$ where $s \in C$ and $s' \in C'$, and (ii) $\Delta(C, a)(C') = \sum_{t' \in C'} \Delta'(s, a)(t)$ where $s \in C$. The following theorem states that quotients preserve the LimAvg-Safety values of an MDP.

**Theorem 6.** *Given an MDP $\mathcal{G}$, a quantitative probabilistic bisimilarity relation $\equiv$, and a limit-average safety objective $f$, the values in $\mathcal{G}$ and $\mathcal{G}/_\equiv$ coincide.*

*Proof.* For every Player 2 strategy $\tau$ in $\mathcal{G}$, we define a Player 2 strategy ($\tau/_\equiv$) in $\mathcal{G}/_\equiv$ (or vice-versa) where: $\tau((s_1, a_1)(s_2, a_2) \ldots (s_n, a_n) \cdot s_{n+1}) = (\tau/_\equiv)((C_1, a_1)(C_2, a_2) \ldots (C_n, a_n) \cdot C_{n+1})$ where $C_i$ is the equivalence class containing $s_i$. By the properties of $\equiv$, it is simple to check that both $\tau$ and $\tau/_\equiv$ have equal values. □

Consider a standard abstraction technique, *data abstraction*, which erases the value of given variables. We show that under certain syntactic restrictions (namely, that the abstracted variables do not appear in any guard statements), the equivalence relation given by the abstraction is a quantitative probabilistic bisimilarity relation and thus is a sound abstraction with respect to any limit-average safety objective. We also consider a less coarse abstraction, *equality and order abstraction*, which preserves equality and order relations among given variables. This abstraction defines a quantitative probabilistic bisimilarity relation under the syntactic condition that the guards test only for these relations, and no arithmetic is used on the abstracted variables.

## 5 Experiments

We describe the results obtained by applying our prototype implementation of techniques described above on four examples. In the examples, obtaining a correct program is not difficult and we focus on the synthesis of optimal programs.

The partial programs were manually abstracted (using the data and order abstractions) and translated into PROMELA, the input language of the SPIN model checker [11]. The abstraction step was straightforward and could be automated. The transition graphs were generated using SPIN. Then, our tool constructed the game graph by taking the product with the scheduler and performance model. The resulting game was solved for the LimAvg-Safety objectives using techniques from Section 4. The examples we considered were small (each thread running a procedure with 15 to 20 lines of code). The synthesis time was under a minute for all but one case (Example 2 with larger values of $n$), where it was under five minutes. The experiments were run on a dual-core 2.5Ghz machine with 2GB of RAM. For all examples, the tool reports normalized performance metrics where higher values indicate better performance.

| LC: CC | Granularity | Performance |
|---|---|---|
| 1:100 | Coarse | 1 |
|  | Medium | 1.15 |
|  | Fine | 1.19 |
| 1:20 | Coarse | 1 |
|  | Medium | 1.14 |
|  | Fine | 1.15 |
| 1:10 | Coarse | 1 |
|  | Medium | 1.12 |
|  | Fine | 1.12 |
| 1:2 | Coarse | 1 |
|  | Medium | 1.03 |
|  | Fine | 0.92 |
| 1:1 | Coarse | 1 |
|  | Medium | 0.96 |
|  | Fine | 0.80 |

**Table 1.** Performance of shared buffers under various locking strategies: LC and CC are the locking cost and data copying cost

**Example 1.** We consider the producer-consumer example described in Section 1, with two consumer and two producer threads. The partial program models a four slot concurrent buffer which is operated on by producers and consumers. Here, we try to synthesize lock granularity. The synthesis results are presented in Table 1.

The most important parameters in the performance model are the cost of locking/unlocking $l$ and the cost $c$ of copying data from/to shared memory. If $c$ was higher than $l$ (by 100:1), then the fine-grained locking approach is better (by 19 percent), and is the result of synthesis. If the cost $l$ is equal to $c$, then the coarse-grained locking approach was found to perform better (by 25 percent), and thus the coarse-grained program is the result of the synthesis.

**Example 2.** We consider the optimistic concurrency example described in detail in Section 1. In the code (Figure 1), the number of operations performed optimistically is controlled by the variable n. We synthesized the optimal n for various performance models and the results are summarized in Table 2.

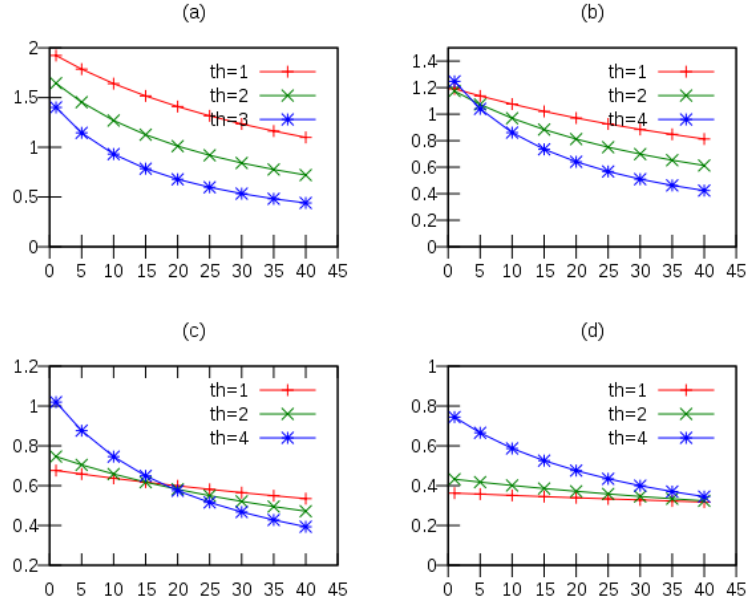

**Fig. 5.** Work sharing for initialization costs and thread counts: Increasing amount of work is shared for the cases (a), (b), (c) and (d)

We were able to find correspondence between our models and the program behavior on a desktop machine: (a) We observed that the graph of performance-vs-`n` has a local maximum when we tested the partial program on the desktop. In our experiments, we were able to find parameters for the performance model which have similar performance-vs-`n` curves. (b) Furthermore, by changing the cost of locking operations on a desktop, by introducing small delays during locks, we were able to observe performance results similar to those produced by other performance model parameters.

**Example 3.** We synthesize the optimal number of threads for work sharing (pseudocode in Figure 6). For independent operations, multiple threads utilize multiple processors more efficiently. However, for small number of operations, thread initialization cost will possibly overcome any performance gain.

| WC : LC | LWO | Performance for n ||||| 
|---|---|---|---|---|---|---|
|  |  | 1 | 2 | 3 | 4 | 5 |
| 20:1 | 1 | 1.0 | 1.049 | 1.052 | 1.048 | 1.043 |
| 20:1 | 2 | 1.0 | 0.999 | 0.990 | 0.982 | 0.976 |
| 10:1 | 1 | 1.0 | 1.134 | 1.172 | 1.187 | 1.193 |
| 10:1 | 2 | 1.0 | 1.046 | 1.054 | 1.054 | 1.052 |

**Table 2.** Optimistic performance: WC, CC, and LWO are the work cost, lock cost, and the length of the work operation

The experimental results are summarized in Figure 5. The x- and y- axes measure the initialization cost and performance, respectively. Each plot in the graph is for a different number of threads. The four graphs (a), (b), (c), and (d) are for a different amounts of work to be shared (the length of the array to be operated was varied between 8, 16, 32, and 64).

```
main:
  n = choice(1..10);
  i = 0;
  array[0..N];
  while (i < n) {
    spawn(worker, i * (N/n), (N/n));
    i++;
  }

worker(start, length):
  i = start;
  while(i < start + length) {
    work(array[i]);
  }
```

**Fig. 6.** Pseudo-code for Example 3

As it can be seen from the figure, for smaller amounts of work, spawning fewer threads is usually better. However, for larger amounts of work, greater number of threads outperforms smaller number of threads, even in the presence of higher initialization costs. The code was run on a desktop (with scaled parameters) and similar results were observed.

**Example 4.** We study the effects of processor caches on performance using a simple performance model for caches. A cache line is modeled as in Figure 6. It assigns differing costs to read and write actions if the line is cached or not. The performance model is the synchronous product of one such automata per memory line. The only actions in the performance model after the synchronous product (caches synchronize on `evict` and `flush`) are `READ` and `WRITE` actions. These actions are matched with the transitions of the partial program.

The partial program is a pessimistic variant of Figure 1 (pseudocode shown in Figure 7). Increasing n, i.e., the number of operations performed under locks, increases the temporal locality of memory accesses and hence, increase in performance is expected. We observed the expected results in our experiments. For instance, increasing n from 1 to 5 increases the performance by a factor of 2.32 and increasing n from to 10 gives an additional boost of about 20%. The result of the synthesis is the program with $n = 10$.

```
1: while(true) {
2:   n  = choice(1..10);
3:   lock();
4:   while (i < n) {
5:     data = write(work(
              read(data)));
6:   }
7:   unlock(lock);
8:}
```

**Fig. 7.** Pseudo-code for Example 4

## 6   Conclusion

**Summary.** Our main contributions are: (1) we developed a technique for synthesizing concurrent programs that are both correct and *optimal*; (2) we introduced a parametric performance model providing a flexible framework for specifying performance characteristics of architectures; (3) we showed how to apply imperfect-information games to the synthesis of concurrent programs and established the complexity for the game problems that arise in this context (4) we developed and implemented practical techniques to efficiently solve partial-program synthesis, and we applied the resulting prototype tool to several examples that illustrate common patterns in concurrent programming.

**Future work.** Our approach examines every correct strategy. There is thus the question whether there exists a practical algorithm that overcomes this limitation. Also, we did not consider the question which solution(s) to present to the programmer in case there is a number of correct strategies with the same performance. Furthermore, one could perhaps incorporate some information on the expected workload to the performance model. There are several other future research directions: one is to consider the synthesis of programs that access concurrent data structures; another is to create benchmarks from which performance automata can be obtained automatically.

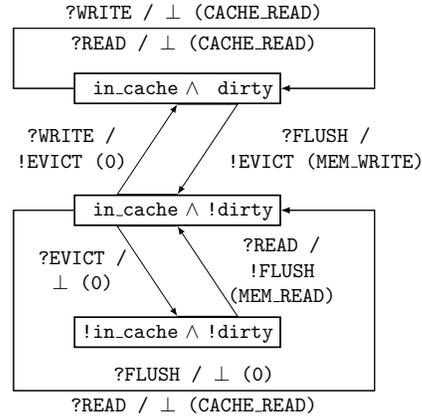

**Fig. 8.** Perf. aut. for Example 4